\documentclass[12pt]{article} 

\usepackage{mathtools, amsfonts, amsthm, latexsym, amssymb,dsfont}
\usepackage[T1]{fontenc}
\usepackage[tracking=true]{microtype}
\usepackage{lmodern}
\usepackage{verbatim}
\usepackage[margin=1in]{geometry}

\usepackage{verbatim}
\usepackage{graphicx}
\usepackage{subcaption}
\usepackage{booktabs}
\usepackage{longtable}
\usepackage{cite}

\usepackage{color}
\definecolor{darkblue}{cmyk}{0.9,0.9,0,0}
\definecolor{wine-stain}{rgb}{0.5,0,0}
\usepackage[colorlinks, allcolors=darkblue, linktoc=all]{hyperref} 
\usepackage{setspace}

\renewcommand{\L}{\mathcal L}

\newcommand{\beq}{\begin{equation}}
\newcommand{\eeq}{\end{equation}}
\newcommand{\be}{\begin{equation}}
\newcommand{\ee}{\end{equation}}
\newcommand{\bea}{\begin{eqnarray}}
\newcommand{\eea}[1]{\label{#1}\end{eqnarray}}

\newcommand{\mn}{\mu\nu}

\def\g {{\gamma}}
\def\beq{\begin{eqnarray}}
\def\eeq{\end{eqnarray}}

\def\p{{\cal P}}

\def\g{{\cal G}}

\def\L*{{\cal L}_*}
\def\L{\mathcal{L}}
\def\({\left(}
\def\){\right)}

\def\p{\partial}

\def\p{\partial}

\def\<{\langle}
\def\>{\rangle}

\def\g{\bar g}
\def\R{\bar R}

\def\M{\bar M}
\def\n{\nabla}

\def\xyma{\xymatrix@M.7em}
\def\xymas{\xymatrix@M.1em}
\newcommand{\ba}{\begin{eqnarray}}
\newcommand{\ea}{\end{eqnarray}}

\begin{document}
\thispagestyle{empty}
\vspace*{0.5in} 
\begin{center}
{\Large \bf Holography for the Trace Anomaly Action}
\vskip 0.25cm

\vskip 0.7cm
\centerline{
{\large {Gregory Gabadadze$^a$ and Massimo Porrati$^{a,b}$}}
}
\vspace{.2in} 

\centerline{{\it a) Center for Cosmology and Particle Physics, Department of Physics}}
\centerline{{\it New York University, 726 Broadway, New York, NY, 10003, USA}}

\vspace{.1in}
\centerline{{\it b) The Blackett Laboratory, Imperial College London}}
\centerline{{\it Prince Consort Road, London SW7 2AZ, UK}}

\end{center}

\begin{abstract}

A recently proposed effective action for the trace anomaly 
describes a tensor-scalar theory that is weakly  coupled up to  a certain high energy scale,
where it becomes strongly interacting. Its ultraviolet completion is obtained by coupling to gravity a 
quantum field theory in which conformal invariance is spontaneously broken. 
In this paper, we show that if the field theory that gives rise to the trace 
anomaly is a large $N_c$ conformal field theory, then the trace anomaly action 
has a completion above the strong scale in a holographic Randall-Sundrum 
two-brane theory, with the radion as a low energy remnant 
of the spontaneously broken conformal symmetry. Furthermore, we note that the sub-leading 
$N_c$ terms can be derived by adding localized fields to the UV brane, so 
that the theory remains weakly coupled.  The sub-leading terms are also obtained
by introducing the Weyl squared terms in the 5D bulk. These, however, exhibit  strongly 
coupled behavior at the respective sub-Planckian energy scales.

\end{abstract}

\newpage
\noindent

\section{Introduction and summary}

A local diffeomorphism-invariant action capturing the gravitational trace anomaly \cite {Duff} has been 
know thanks to the works of Riegert, and Efim Fradkin and Tseytlin \cite {Riegert,FT}; it is referred to as the local Riegert 
action, albeit the works \cite {Riegert} and \cite {FT} are practically simultaneous,  with Riegert's 
derivation being more general. The current understanding of the Riegert action as an effective action for the  
trace anomaly was achieved by Komargodski and Schwimmer \cite {KS} (see also an $SO(2,4)/ISO(1,3)$ 
coset construction leading  to this action in \cite {GGTukhashvili})\footnote{After deriving the local but seemingly 
non-covariant anomaly action, Riegert went on to rewrite it as a manifestly diff-invariant, but non-local action \cite {Riegert}. 
This rewriting introduced a fourth-order derivative into the action via a non-local field redefinition used 
in eq. (19) of \cite {Riegert}.  The resulting action in eqs. (24,25) of \cite {Riegert} 
has a ghost because of the fourth-order derivative term. This ghost, even if projected out classically, 
will still lead to unphysical quantum instabilities. For that reason,
the anomaly actions containing fourth-order derivatives will not be considered here. 
In this work we only use Riegert's local action, presented in eqs. (6,8) of \cite {Riegert}, which is diffeomorphism  
invariant \cite {KS}, and has no four-derivative kinetic terms.}. 

However, General Relativity (GR) 
coupled to the Riegert action is a strongly coupled theory at an arbitrary low energy scale (see \cite {Kurt,GG}, and references therein).  
One of us proposed in \cite {GG} to resolve this problem by augmenting  the classical GR action so 
that the amended theory,  together with the Riegert action,  is weakly coupled all the way up to a certain  
high energy scale $\bar M$, which could presumably be in the interval,   
$M_0 \sim10^5\,GeV\ll  {\bar M} \ll  M_{\rm Pl}=1/\sqrt{8\pi G}\sim 10^{18}\, GeV$.
The augmented action  \cite {GG}, without the Riegert term,  reads:
\beq
S_{R-\R}=M^2 \, \int d^4 x \sqrt{-g}\,R\,  -  {\bar M}^2 \, \int d^4 x \sqrt{-{\bar g}}\,{\bar R}\,,
\label{aGR}
\eeq
where ${\bar R}\equiv R(\g)$, $M=M_{\rm Pl}/\sqrt{2}\gg {\bar M}$, and  the 
two metric tensors are related as 
\beq
g_{\mu\nu} = e^{2 \tau} \, {\bar g}_{\mu\nu}\,,
\label{gsgbar}
\eeq
where $ \tau$ is a scalar field.
Consequently,  the total  effective action that captures classical gravitational physics and 
the trace anomaly equation reads as follows \cite {GG}:
\beq
S_{eff}=   S_{R-\R}\, + \, S_{A}( \tau, {\bar g} )\,,
\label{S00}
\eeq
where $S_A$ -- denoting the local Riegert action -- has the form \cite {Riegert,FT,KS} 
\beq
S_{A} = -2 a\int d^4x \sqrt {- {\bar g} } \left (  \tau {\bar E} - 4 {\bar G}^{\mu\nu} {\bar \nabla}_\mu  \tau {\bar \nabla}_\nu  \tau  - 4 ({\bar \nabla}^2  \tau) ({\bar \nabla}  \tau)^2 - 
2 ({\bar \nabla}  \tau )^4\right) + 2 c'\int d^4x \sqrt {- {\bar g}}  \tau {\bar W}^2,
\label{SA0}
\eeq
with the covariant derivative $\bar \n$, the Euler (Gauss-Bonnet) invariant, 
${\bar E}={\bar R}_{\mu\nu\alpha\beta}^2 - 4 {\bar R}_{\mu\nu}^2+{\bar R}^2$, and the Weyl tensor 
squared, ${\bar W}^2={\bar R}_{\mu\nu\alpha\beta}^2 - 2 {\bar R}_{\mu\nu}^2+{\bar R}^2/3$, all made out of the metric $\bar g$.

The action (\ref {S00}), with $c^\prime=0$, was 
introduced earlier in \cite {Fernandes} in a different, entirely classical context 
without  reference to the quantum trace anomaly, but as a scalar-tensor action maintaining two 
derivative conformally invariant equations of motion, in spite of the action itself not being 
conformally invariant.  This action belongs to a more general class of  Horndeski's 
scalar-tensor theories, which maintain second order equations of motion \cite {Horndeski}. 

It is evident that eq.~(\ref {S00}) should be regarded as defining an effective low energy action. 
Since $\bar M \ll M_{\rm Pl}$, all Planck-scale suppressed higher dimensional  terms 
can be ignored in this action; it  describes a  weakly coupled theory up to the scale $\bar M$, 
where it becomes strongly coupled \cite {GG}.

Recall that the GR action breaks the scale invariance explicitly, with $M_{\rm Pl}$ 
being the breaking scale\footnote{To be more precise, the scale transformations 
here refer  to the transformations formed by the dilatations together with simultaneous global diffeomorphisms,  
such that the coordinates do not transform under the combined transformations, 
but the metric tensor transforms as, $g\to e^{2\lambda } g$, with a constant $\lambda$.}. 
The parameter $\bar M$, on the other hand, can 
be viewed as a scale where the  conformal symmetry is spontaneously 
broken \cite {GGTukhashvili,GG}. The low energy remnant 
of the spontaneous breaking is $ \tau$. Thus, the action  (\ref {S00}), when viewed as a functional of 
$g$ and $\tau$, contains terms for explicit, spontaneous, and anomalous breaking of 
the conformal symmetry. That being said, the respective high energy theory above $\bar M$ is 
likely to contain more terms and degrees of freedom. The goal of this paper is to discuss the
UV completion of the above action at the scale  $\M$. 

The Riegert action is in fact part of the universal action describing the dynamics of a 
Conformal Field Theory (CFT) with spontaneously broken conformal 
invariance below the breaking scale~\cite{KS}. The field $ \tau$ is identified with the dilaton, 
that is the Nambu-Goldstone boson of spontaneously 
broken conformal invariance,  and the second term in~(\ref{aGR}) contains its 
kinetic term. 

 When a generic CFT is coupled to gravity many relevant operators -- such as scalar masses -- are 
induced, which make the theory gapped and lift the flat directions in the scalar potential. The flat directions are the very 
reason why conformal invariance is broken spontaneously instead of being broken explicitly, 
so spontaneous conformal symmetry breaking is nongeneric in quantum field theory. An exception to this scenario is 
represented by theories that come from certain highly
supersymmetric compactifications of string theory. 

Another interesting example is large-$N_c$ CFTs, which 
have a holographically dual description in 5D.
Here we  will study the UV completion of the action~(\ref{S00}) in the case when a 
Quantum Field Theory (QFT) that gives rise to the Riegert action is a large $N_c$ CFT  that has 
a holographic dual. There are simplifications in this case that help our analysis. Indeed, 
the general trace anomaly equation, $ \langle T^\mu_\mu \rangle = -aE+c'W^2$,
simplifies in the large $N_c$ limit of a CFT  where $a=c'$ 
\beq
\langle T^\mu_\mu \rangle = a\,(W^2-E )= {N_c^2 \over 32 \pi^2} \left ( R_{\mu\nu}^2 -{1\over 3} R^2 \right )\,.
\label{trace}
\eeq
 Henningson and Skenderis have shown in \cite {HenningsonSkenderis} that 
the term, $R_{\mu\nu}^2 -{1\over 3} R^2  $, generically emerges in equations of motion 
via 5D $AdS$ holography, where  they derived the trace anomaly equation 
for a large $N_c$ CFT (\ref {trace}). One would therefore expect 
that the corresponding 4D effective action for the trace anomaly could  
also be  derived from the 5D $AdS$ theory. We will show how this 
expectation is  indeed fulfilled. 

 After writing  5D gravity in the formalism that we will use throughout the paper (see, Section~\ref{5D})
 we will derive the Riegert action as an effective action in Section~\ref{r5d}. The 4D action would 
 appear non-local if only the metric field were to be used. However, it can be rewritten as a local 
 difff-invariant action by integrating into it the  field $ \tau$. Since $ \tau$ is also the dilaton, we need a 5D 
 model for spontaneously broken conformal invariance coupled to gravity. This requirement 
 can be realized by  the two-brane Randall-Sundrum model~\cite{rs1} (RS1), which is 
 indeed holographic to a CFT coupled to gravity~\cite{ahpr}. Importantly, the position of the 
 IR brane in RS1, which is not stabilized,  is precisely the massless dilaton, as shown in~\cite{RZ}. 
  
 In Section~\ref{4d} we will  recover the result of~\cite{RZ,cgrt,gw} that 
 shows  how  the $R-{\bar R}$ term naturally emerges in the 
 holographic picture of the RS1 model. The main result of Section 4 is to show that 
 the RS1 action also generates the Riegert term with $a=c'$, together with certain 
 specific conformally-invariant dimension-4 terms. Unlike the Riegert terms, the  
 latter are non-universal and model-dependent.  In Section~\ref{holo} we show that our action satisfies the 
 Ward identities of spontaneously broken conformal invariance. In Section~\ref{a-c} we will study deviations
 from the $a=c'$ limit. In Section \ref {disc} we will discuss some open questions.

\section{The 5D theory}
\label{5D}

First of all we summarize  the 5D action and equations in the  formalism developed  
by Shiromizu, Maeda, and Sasaki \cite {ShiromizuMaedaSasaki}. The bulk action is
\beq
S_{\rm Bulk} = M_5^3 \,\int d^4x dz \sqrt {\hat g} N \left ( {\hat R} +{\hat K}^2 - {\hat K}_{\alpha\beta}^2 - 2 \Lambda_5 \right),
\label{5Daction}
\eeq
where $N$ is the lapse in the 5th dimension, and $N_\mu$ is the respective shift 
(see ref.~\cite{DeffayetMourad} for more details). The 5D cosmological constant will be chosen to be negative,
$-2\Lambda_5 =12/L^2$, where $L$ is the radius-curvature of a 5D $AdS$ spacetime. 
The extrinsic curvature of 4D spacetime  is defined as follows
\beq
{\hat K}_{\mu\nu} = {1\over 2N} \left (\partial_z  {\hat g}_{\mu\nu} - {\hat \nabla}_\mu N_\nu -  {\hat \nabla}_\nu N_\mu \right ).
\eeq
Variation of the bulk action (\ref {5Daction}) w.r.t. the lapse $N$ gives
\beq
{\hat R} - 2\Lambda_5 = {\hat K}^2 -{\hat K}_{\alpha\beta}^2,
\label{55}
\eeq
and the equation obtained by variation of the action w.r.t. $N_\beta$  reads
\beq
{\hat \nabla}^\alpha {\hat K}_{\alpha\beta} = {\hat \nabla}_\beta {\hat K}.
\label{mu5}
\eeq
As long as the above equation is satisfied, we can substitute $N_\mu=0$
in the bulk $\{\mu\nu\}$ equation, which  then reads as follows:
\beq
{\hat G}_{\mu\nu} + \Lambda_5 {\hat g}_{\mu\nu} = {1\over 2}  {\hat g}_{\mu\nu} 
\left (  {\hat K}^2 -{\hat K}_{\alpha\beta}^2    \right ) + 2 \left (     {\hat K}_\mu^\rho {\hat K}_{\rho \nu}  -{\hat K} {\hat K}_{\mu\nu} \right ) \nonumber  \\
+\,{{\hat \nabla}_\mu {\hat \nabla}_\nu N - {\hat g}_{\mu\nu} {\hat \nabla}^2 N \over N} -  {\hat g}_{\mu\alpha} {\hat g}_{\nu\beta}
 {  \partial_z \left (\sqrt{-{\hat g}} ({\hat K} {\hat g}^{\alpha\beta} - {\hat K}^{\alpha\beta} ) \right  )  \over  N \sqrt{-{\hat g}} }.
\label{munu}
\eeq
We will use these equations below, together with the appropriate  boundary conditions, 
to determine an effective 4D theory. 

In the next section  we will consider only one brane, together with the corresponding boundary conditions specified there. 
We will show that by integrating out the bulk in the single-brane RS model one gets at low energies 
4D GR plus the  Riegert action, with $a=c'$.

\section{The Riegert action from $AdS_5$}\label{r5d}

After taking into account the equations  of motion, (\ref {55}) and (\ref {mu5}),
we can set $N_\mu=0$,  $N=A(z)=L/(z+L)$   for $z\ge 0$, 
and consider the metric:
\beq
ds^2 = {\hat g}_{\mn}(x,z) dx^\mu dx^\nu + A^2(z) dz^2\,.
\label{metric} 
\eeq
We will follow Kanno and Soda \cite {KannoSoda1} to integrate out  the 5D bulk.   
This is done  via a classical nonlinear order-by-order expansion 
of the 5D equations of motion in powers of $RL^2$, where $R$ is a 4D curvature
experienced by a 4D observer on the positive tension brane in the single-brane RS model. 
The corresponding expansion of the metric is parametrized as follows:
\beq
{\hat g}_{\mn}(x,z) = A^2(z) \(g_{\mn}(x) +g^{(1)}_{\mn}(x,z) +g^{(2)}_{\mn}(x,z)+...   \),
\label{gexp}
\eeq
where $g^{(j)}_{\mn} (x,z=0)=0,\, {\rm for}\,j=1,2,3,..$. Using the above expansion one can find
the  corresponding power series expression  for the extrinsic curvature
\beq
{\hat K}^\mu_\nu = \sum^{\infty}_{n=0} K^{(n)\mu}_\nu \,.
\label{Kexp}
\eeq
The expression for $K^{(2)\mu}_\nu $ depends of an unknown tensor that 
can't be written as a variation of a local tensor made out of the metric and its derivatives \cite {KannoSoda1}.  
On the other hand, traces of $K^{(n)\mu}_\nu $  can be determined unambiguously  in terms of local 
quantities \cite {KannoSoda1}. We use this observation  and come up with a scheme 
that only utilizes  the traces of the extrinsic curvatures evaluated at $z=0$:
\beq
K^{(0) \mu}_\mu|_{z=0} = -{4\over L}, ~~~K^{(1)\mu}_\mu |_{z=0} = -{L \over 6} R(g),~~~K^{(2)\mu}_\mu 
|_{z=0} =-{L^3 \over 24} \left (R_{\mu\nu}^2 -{1\over 3}R^2 \right ).
\label{Kexp2}
\eeq
Taking the trace of the junction condition at $z=0^+$ 
\beq
T_4 {\hat g}_{\mu\nu}  -  2M_5^3 
\left (  {\hat K}_{\mu\nu} -  {\hat K} {g}_{\mu\nu}   \right ) =0\,,
\label{BC1}
\eeq
one gets
\beq
4 T_4+ 6 M_5^3 {\hat K} =0.
\label{BC2}
\eeq
Using the RS fine tuning condition 
\beq
T_4 = 6 M_5^3/L\,,
\label{4cc5cc}
\eeq
and noticing that ${\hat R ({\hat g} )}|_{z=0^+} = R(g)$,  we  can rewrite the trace equation (\ref {BC2})  
as follows:
\beq
M^2 R =  -{M_5^3 L^3 \over 4} \(  R_{\mu\nu}^2 -{1\over 3} R^2 \) \equiv -a (W^2-E)\,,
\label{anom0}
\eeq
where $M^2 = M_5^3 L$ is the  4D Planck mass squared divided by 2, 
and $a\equiv {M_5^3 L^3 \over 8}\sim { N_c}^2$. The above  is the trace anomaly 
equation. As noted earlier, the emergence of the 4D trace anomaly equation from a 5D $AdS$ bulk
was first shown  in \cite {HenningsonSkenderis}.

Our goal is  to derive the action that gives rise to the equation (\ref {anom0}). To achieve this we can follow  
the method used by Riegert \cite {Riegert}. We multiply both sides of the equation (\ref {anom0}) by $\sqrt{-g}$, and perform 
the following field redefinition, $g_{\mu\nu}(x)= e^{2 \tau} {\bar g}_{\mu\nu}$; this gives  
\beq
M^2   \sqrt{-{\bar g}}  e^{2\tau}   \(    {\bar R}({\g}) -6\, ({\bar \n}^2 \tau) - 6 {\bar \nabla}^2 \tau \)   = \nonumber \\
-2a \sqrt {- {\bar g} } \left ( - {{\bar E}\over 2} + {{\bar W}^2\over 2}  - 4 {\bar R}^{\mu\nu} 
( {\bar \nabla}_\mu {\bar \nabla}_\nu  \tau - {\bar \nabla}_\mu  \tau {\bar \nabla}_\nu  \tau)  + 2 {\bar R} {\bar \nabla}^2  \tau \right )
\nonumber \\
 -2 a \sqrt {- {\bar g} } \left ( 
-4 ( {\bar \nabla}^2 \tau)^2 + 4 ( {\bar \nabla}_\mu {\bar \nabla }_\nu  \tau)^2 
- 4({\bar \nabla}^2  \tau) ({\bar \nabla}  \tau)^2 -  8 ({\bar \nabla}^{\alpha} {\bar \nabla}^{\beta} \tau) ({\bar \nabla}_\alpha   \tau) {\bar \nabla}_\alpha   \tau) \right )\,.
\label{anom}
\eeq
We now look for an action, as a functional of $ \tau$ and ${\bar g}$ and their derivatives, 
 that can be varied w.r.t. $ \tau$ to give rise  to 
(\ref {anom}). This action reads
\beq
S=M^2 \, \int d^4 x \sqrt{-{\g}}\,  e^{2\tau}   \(    {\bar R}({\g}) +6\, ({\bar \n} \tau)^2  \)+  I({\bar g}) \nonumber \\ 
-2 a \int d^4x \sqrt {- {\bar g} } \left ( \tau  {\bar E} - 4 {\bar G}^{\mu\nu} {\bar \nabla}_\mu  \tau {\bar \nabla}_\nu  \tau  - 4 ({\bar \nabla}^2  \tau) ({\bar \nabla}  \tau)^2 - 
2 ({\bar \nabla}  \tau )^4\right) +2a \int d^4x \sqrt {- {\bar g}} \, \tau\, {\bar W}^2\,.
\label{actionI}
\eeq
The second line is the Riegert action with $a=c'= M_5^3 L^3/8$, 
consistent with a large $N_c$ CFT.  The first term in the first line is the Einstein-Hilbert 
term written in the Jordan frame. The second term in the first 
line, denoted by $I({\bar g})$, is a functional that does not depend on $ \tau$, but can depend on 
${\bar g}$ and its derivatives. In a sense, $I({\bar g})$ is an "integration functional" which cannot 
be determined by the above procedure. If one assumes $I({\bar g}) =0$, as it was done by Riegert \cite {Riegert}, 
then the theory is strongly coupled at arbitrarily low energies \cite {Kurt,GG}. One way to avoid this problem is to postulate 
that, $I({\bar g})=-{\M}^2 \int d^4x \sqrt{-{\bar g}} {\bar R}$, based on symmetry and anomaly considerations, 
as was done in  \cite {GG}.  

From the 5D perspective the  strong coupling issue stems from the fact that the
gravitational  Kaluza-Klein modes are infinitely strongly coupled at the nonlinear level 
near the $AdS$ horizon in the single-brane RS model \cite {rs2}. This 
issue gets resolved by introducing a second brane which cuts off the $AdS$ horizon in the RS model.
Interestingly,  the expression for $I({\bar g})$ -- precisely  with the negative sign -- naturally 
emerges in the holographic picture as soon as one introduces the second brane in the RS model \cite {rs1}; 
this leads to a radion field, which can then be identified with the $\tau$ field introduced above.  
All this is discussed in gradually increasing detail in the subsequent sections.

\section{Derivation of $I({\bar g}) $}\label{4d}

The expression for $I({\bar g}) $ could be obtained 
from the comprehensive work of Kanno and Soda \cite {KannoSoda2}; 
it is for sake of presentation that 
we will give a different derivation of $I({\bar g}) $ here. 

First we use the bulk equation  (\ref {55}) in the bulk action (\ref {5Daction}) to get a partially "on-shell"  
action 
\beq
S_{\rm Bulk}|  = 2 M_5^3\,\int d^4 x dz \sqrt {\hat g} N \left ( {\hat R} - 2 \Lambda_5 \right).
\label{5DactionN}
\eeq
To introduce a radion field we parametrize the metric as follows:
\beq
ds^2 = {\hat g}_{\mu\nu}(x,z) dx^\mu dx^\nu + N^2(x,z) dz^2\,,
\label{metric2} 
\eeq
and adopt new notations, as well as more general expansion for 
the metric than the one used in the previous section:
\beq
{\hat g}_{\mu\nu}(x,z) \equiv\Omega^2(x,z) \, g_{\mu\nu}(x,z) \equiv \Omega^2(x,z) \(g_{\mu\nu}(x) +\delta g_{\mu\nu}(x,z)    \),
\label{Gexp}
\eeq
where $\delta g_{\mn} (x,z=0)=0$. The metric $g_{\mn}(x)$ denotes the RS zero mode, while 
$\delta g_{\mn}$ encodes sub-leading curvature corrections due to the bulk. 
Using (\ref {metric2},\ref {Gexp})  one can find the corresponding expression  for the extrinsic curvature 
\beq
{\hat K}_{\alpha\beta} = {\partial_z \Omega^2 \over 2 N \Omega^2} \,{\hat g}_{\alpha\beta} (x,z) + { \Omega^2 \over 2 N}
\p_z \delta g_{\alpha\beta}\,.
\label{hatK2}
\eeq
Note that the first term on the r.h.s. contains the leading and subleading pieces in $\hat g$.
Substituting the latter expression into (\ref {mu5}), and focusing on the leading order, 
we get the following   relation between $N$ and $\Omega$ 
\beq
N (x,z) = { U\over 2} {\partial_z \Omega^2 \over \Omega^2}\,,
\label{N}
\eeq
where $U$ is an integration constant.  By substituting (\ref {N})  into the bulk equation (\ref {55})   one  
can determine the integration constant from the obtained relation
\beq
-2 \Lambda_5 = {12 \over U^2} = {12\over L^2}\,. 
\label{csol}
\eeq
We can now use (\ref {N}, \ref {csol}) in (\ref {5DactionN}) to  obtain
\beq
S_{\rm Bulk}|  = {M_5^3 L}\, \int d^4 x \int dz \sqrt {-g}  \left (  \partial_z ( \Omega^2)  R(g) + 6 \partial_z (\n \Omega)^2    -  
(\partial_z \Omega^4 ) \Lambda_5  \right).
\label{5DactionNNmu}
\eeq
It is clear that the $z$ integration can be done explicitly and that the result of that integration depends only 
on the boundary values of $\Omega$.  We choose the following boundary conditions 
\beq
\Omega(x,z=0)=1, ~~\Omega(x,z= \chi (x)) =  \Phi (x) = {L\over z_{IR}+L}\, e^{ -\tau (x)}\,, 
\label{BCOmega}
\eeq
where we parametrized the $x$-dependent boundary, $\chi (x)$, by the  field $\Phi(x)$.
Using the junction conditions at both branes to integrate (\ref {5DactionNNmu}) w.r.t. $z$ 
from the UV, $z=0$, to the IR, $z= \chi (x)$, we obtain 
\beq
M^2 \int d^4 x \sqrt {-g}  \left (R - \Phi^2 R - 6  (\nabla \Phi)^2  \right)= 
M^2 \int d^4 x \( \sqrt {-g} R  - \epsilon^2 \sqrt{-{\bar g}} {\bar R} \right)\,. 
\label{int}
\eeq
Here  $\epsilon = L/(z_{IR}+L)\equiv {\bar M/ M}$, and hence 
we find that $I({\bar g}) =-{\M}^2 \int d^4x \sqrt{-{\bar g}} {\bar R}$.

Note that we require ${\bar M}\ll M$. This hierarchy 
can be achieved if  the average distance between the two branes is much 
greater than the radius curvature of $AdS_5$,  $z_{IR} \gg L$. This implies that that the 
mass scale of the lightest Kaluza-Klein (KK) modes, $\sim z_{IR}^{-1}$,  is below the scale of the 
curvature of $AdS_5$.   Moreover,
we find that, ${\bar M} \sim N_c/z_{IR}$, which is much higher than the KK mass scale, $z_{IR}^{-1}$, 
and, by construction, is  much smaller than the 4D Planck scale, $M_{\rm Pl}\sim M\sim N_c/L$.

The action (\ref{int}) was found in~\cite{cgrt,gw,RZ} by dimensional reduction of the 
5D RS1 action. We will also derive it in the next sections by a dimensional 
reduction that solves the radial Hamiltonian and momentum constraints 
eqs.~(\ref{55},\ref{mu5}). By taking care of the constraints we
can also compute the first order correction to the action, which depends 
on $\delta g_{\mu\nu}$. This can be done systematically
using the formalism of ref.~\cite{KannoSoda2} and by 
selecting appropriate integration constants in their equations. When considering
the first correction to the lowest-order results -- i.e. $\delta g_{\mu\nu}=g^{(1)}_{\mu\nu}\equiv h_{\mu\nu}$ in the expansion~(\ref{Gexp}) -- we can simplify the derivation and explicitly perform integrals in 
$z$ that are left implicit in~\cite{KannoSoda1,KannoSoda2}. 
We can also find at the same time the Riegert action and the 
next to leading Weyl-invariant terms.

\subsection{Solving the equations for the extrinsic curvature $K_{\mn}$}

We are interested in an effective field theory in which the expansion parameter is $z_{IR}$ times gradients of the fields. 
The first term in
the expansion is eq.~(\ref{int}) while the second is obtained by expanding $g_{\mn}(x,y)= g_{\mn}(x) + h_{\mn}(x,y)$ in the 
action~(\ref{5Daction}) up to the fourth order in  $z_{IR} \partial_\mu$. It is also convenient
to define a new radial coordinate $y$ as $z=z_{IR} y -L$, and since $0\le z\le z_{IR}$ in this section, 
the integration range for $y$ is 
 ${L\over z_{IR}}\leq y \leq 1+{L\over z_{IR}} $. In the rest of this section and in Section 6 we also use 
 a different definition of $\Omega$, which is  $ {(z+L)/L}$ times the $\Omega$ used in the previous section. With this new definition the exact Anti de Sitter metric $ds^2 ={L^2\over (L+z)^2}( \eta_{\mu\nu} dx^\mu dx^\nu + dz^2)$ corresponds to $\Omega=1$. We find
\bea
S_{Bulk}&=& {c\over z_{IR}^2} \int_{L/z_{IR}}^{1+L/z_{IR}} {dy\over y^3} \int d^ 4 x  \sqrt{-g} \Big[N\Omega^2 (\n_\mu\n_\nu h^{\mn}-\n^2 h 
- h^{\mn} G_{\mn}) \nonumber \\ &&-6 h^{\mn}\n_\mu(N\Omega) \n_\nu \Omega +3 h \n_\mu (N\Omega) \n^\mu \Omega  +
 {\Omega^4\over 4N z_{IR}^2} \left( {dh \over dy}{dh\over dy}- {dh^{\mn} \over dy}{d h_{\mn}\over dy}  \right) \Big] . \nonumber \\ &&
\eea{25}
Here, $c=M_5^3 L^3 $,  $h^{\mn}\equiv g^{\mu\rho}g^{\nu\sigma} h_{\rho\sigma}$, $h\equiv g^{\mn}h_{\mn}$, $G_{\mn} =R_{\mn}-{1\over 2} g_{\mn} R$ and
$K_{\mn}= {1\over 2N} {d\over dy} h_{\mn}$, since we set $N_\mu=0$. The equations of motion for $h_{\mn}$ are
\bea
{1\over z_{IR}^2}{d\over dy}\left[ {\Omega^4\over 2 N y^3} \left({d h_{\mn}\over dy} - g_{\mn}{dh \over dy}\right)\right] &= & 
{1\over y^3} \Big( g_{\mn}\n^2 (N\Omega^2)
-\n_\mu \n_\nu (N \Omega^2) +
G_{\mn} \Omega^2 +  \nonumber \\ && 3\n_\mu(N\Omega)\n_\nu \Omega  +3\n_\nu(N\Omega)\n_\mu \Omega
-3g_{\mn} \n_\lambda(N\Omega) \n^\lambda\Omega \Big) \nonumber . \\  &&
\eea{26}
Using the identities
\beq
g^{\mn}\n_\mu(N\Omega)  \n_\nu  \Omega= -{1\over 2} y^3 {d\over dy} (y^{-2} g^{\mn} \n_\mu \Omega \n_\nu \Omega) ,
\quad 
N \Omega^2 = -{1\over 2} y^3 {d\over dy} (y^{-2} \Omega^2)  ,
\label{20}
\eeq
we find that the l.h.s. of~(\ref{26}) is a total derivative so we can write
\bea
    {\Omega^4\over  z_{IR}^2 N y^3} \left({d h_{\mn}\over dy} - g_{\mn}{dh \over dy}\right) &=& {1\over y^2}\Omega^{2} X_{\mn} + C_{\mn}, \nonumber \\
X_{\mn} &\equiv & \Omega^{-2} \big( \n_\mu\n_\nu \Omega^2
-g_{\mn} \n^2 \Omega^2 -G_{\mn} \Omega^2 +\nonumber \\ 
&& -6\n_\mu \Omega \n_\nu \Omega + 3g_{\mn} \n_\lambda \Omega \n^\lambda \Omega \big)  .
\eea{27}
The $y$-independent integration function $C_{\mn}$ can be constrained by demanding that~(\ref{27}) solves the momentum
constraint~(\ref{mu5}).  Using the Bianchi identity $\n^\mu G_{\mn}=0$ and the identity
\beq
(\n_\mu \n_\nu \n^\mu  -\n_\nu \n^2)S = R_{\mn}\n^\mu S ,
\label{28}
\eeq
which is valid for any scalar $S$, we find
\beq
\n^\mu\left[ {\Omega^4\over  z_{IR}^2 N y^3} \left({d h_{\mn}\over dy} - g_{\mn}{dh \over dy}\right)\right] =
{1\over y^2} \partial_\mu \Omega ( \Omega R -6 \n^2\Omega) +\n^\mu C_{\mn}.
\label{29}
\eeq
This solves the constraint  if $\n^\mu C_{\mn}=0$. 

Furthermore, it is convenient to rewrite $X_{\mn}$ in terms
of a new variable $\sigma$ defined as $\Omega=\exp(\sigma)$. We find:
\beq 
X_{\mn}= -2\partial_\mu \sigma \partial_\nu \sigma  + 2 \n_\mu\partial_\nu \sigma -G_{\mn}  -g_{\mn} \partial_\lambda \sigma \n^\lambda\sigma
-2 g_{\mn} \n^2 \sigma.
\label{30}
\eeq
Another useful identity is
$X_{\mn}=-G_{\mn}(\Omega^2 g)=-G_{\mn}(e^{2\sigma} g)$.

\subsection{The boundary condition at $y=1$}
We need to find the boundary conditions for $K_{\mn}$ at $y=1+L/z_{IR}\approx 1$ (recall that $L/z_{IR} \ll 1$). The variation of eq.~(\ref{25}) contains the boundary term
\bea
&&{c\over z_{IR}^2} \int d^4x \sqrt{-g} {\Omega^4\over 2 L^2 N} \left({d h_{\mn}\over dy} - g_{\mn}{dh \over dy}\right)\delta h^{\mn}\Big|_{y=1}
= \nonumber \\ && -c\int d^4 x \sqrt{-g} \Big[e^{-2\tau} G_{\mn}(e^{-2\tau} g) -C_{\mn}\Big] \delta h^{\mn}\Big|_{y=1} ,
\eea{31}
where $\tau\equiv -\sigma |_{y=1}$.
Note that  there is no boundary term at $y=L/z_{IR}$ because there we impose the Dirichlet boundary conditions $\delta h_{\mn}=0$.
We can decompose,  $\delta h_{\mn}=\delta h_{\mn}^{TT} +\n_\mu \xi_\nu + \n_\nu \xi_\mu -2 g_{\mn} \omega$, and notice that the gradient and trace term in the variation can be canceled by varying $\psi$ in the action of the zero modes~(\ref{int}) 
according to $\delta \psi =\xi^m\n_\mu\psi + \omega \psi$. So, to make the action stationary we have to  
impose only that the transverse-traceless part of $K_{\mn}$ vanish: $K^{TT}_{\mn}=0$. This identifies $C_{\mn}$
as $C_{\mn}=\big[ e^{-2\tau} G_{\mn}(e^{-2\tau} g )\big]^{TT}$.

\subsection{The effective action to $O(z_{IR}^0)$}
Let us substitute the solution of the $h_{\mn}$ equations of motion into the action~(\ref{5Daction}). Since we chose free boundary conditions 
at $y=1$,  and the Dirichlet boundary conditions at $y=L/z_{IR}$, we can discard the boundary contributions and get
the action $S_{eff} + S_1$,  with $S_{eff}$ given in eq.~(\ref{int}) and 
\bea
S_1&=&-{c\over {z_{IR}}^2}  \int_{L/z_{IR}}^1 {dy\over y^3} \int d^4 x  \sqrt{-g} 
 {\Omega^4\over 4N {z_{IR}^2} }\left( {dh \over dy}{dh\over dy}- {dh^{\mn} \over dy}{d h_{\mn}\over dy}  \right) \Big]
= \nonumber \\ && 
 {c\over 4} \int_{L/z_{IR}}^1 dy y^3 \int d^4 x  \sqrt{-g} N
 \Omega^{-4} \Big[\Big(  -{1\over y^2} \Omega^2 G_{\mn}(\gamma)) +C_{\mn} \Big)^2 -{1\over 3} \Big({1\over y^2} 
 \Omega^2 g^{\mn}R_{\mn} (\gamma) +C \Big)^2 \Big] . \nonumber \\ &&
   \eea{32}
Here, $\gamma_{\mn} (x,y) \equiv \Omega^2(x,y) g_{\mn}(x)$. A key identity that follows from eq.~(\ref{30}) is
\beq
{d\over dy} G_{\mn}(\gamma)= -2\mathcal{D}_\mu \mathcal{D}_\nu {d\sigma \over dy} + 2\gamma_{\mn} \mathcal{D}^2 {d\sigma\over dy},
\label{33}
\eeq
where $\mathcal{D}_\mu$ denotes the covariant derivative w.r.t. the metric $\gamma_{\mn} (x,y)$.
Expanding equation~(\ref{32}) in powers of $C_{\mn}$ we get three terms.
Using integration by part in $y$ we rewrite the first term as
\bea
&& {c\over 4} \int_{L/z_{IR}}^1 {dy\over y}  \int d^4 x  \sqrt{-g} N \Big(g^{\mu\rho}g^{\nu\sigma}G_{\mn}(\gamma)G_{\rho\sigma}(\gamma)  -{1\over 3}[ g^{\mn}G_{\mn}(\gamma)]^2\Big)  \nonumber \\ &=& 
-{c\over 4} \int_{L/z_{IR}}^1 dy 
\int d^4 x  \sqrt{-g} \Big({d\over dy}(\sigma -\log y)\Big)  \Big(g^{\mu\rho}g^{\nu\sigma}G_{\mn}(\gamma)G_{\rho\sigma}(\gamma)  
-{1\over 3}[ g^{\mn}G_{\mn}(\gamma)]^2\Big)  \nonumber  
\\ &=& -{c\over 4}  \int d^4 x  \sqrt{-g}\Big[ -\tau\Big( G_{\mn}^2(e^{-2\tau} g)  - {1\over 3} G^2(e^{-2\tau} g) \Big) 
+\log(L/z_{IR} ) \Big(G_{\mn}^2(g) - {1\over 3} G^2(g) \Big)\Big]  \nonumber \\ &&
 -c \int_{L/z_{IR}}^1 dy \int d^4 x \sqrt{-\gamma}(\sigma- \log y) \Big(\mathcal{D}_\mu\mathcal{D}_\nu {d\sigma \over dy}
\Big) G_{\rho\sigma}(\gamma)\gamma^{\mu\rho}
\gamma^{\nu\sigma} .
\eea{34}
Notice that we define $G_{\mn}^2= G_{\mn}G_{\rho\sigma} g^{\mu\rho} g^{\nu\sigma}$ etc.
An integration by parts in $\mathcal{D}_\mu$ transforms the last term into
\bea
&& c \int_{L/z_{IR}}^1 dy \int d^4 x \sqrt{-g}\partial_\mu\sigma \left(\partial_\nu {d\sigma \over dy}\right)
 G_{\rho\sigma}(\gamma)g^{\mu\rho}g^{\nu\sigma} \nonumber  \\ &=& 
 {c\over 2} \int_{L/z_{IR}}^1 dy \int d^4 x \sqrt{-g}\left({d\over dy} \partial_\mu\sigma \partial_\nu \sigma \right)
 G_{\rho\sigma}(\gamma)g^{\mu\rho}g^{\nu\sigma} \nonumber \\ &=&
 {c\over 2} \int d^4 x \sqrt{-g} \partial_\mu\tau \partial_\nu \tau
 G^{\mn}(e^{-2\tau} g)
 + c \int_0^1 dy \int d^4 x \sqrt{-\gamma}\gamma^{\mu\rho}\gamma^{\nu\sigma} \partial_\mu\sigma \partial_\nu \sigma 
 \left( \mathcal{D}_\rho \mathcal{D}_\sigma - \gamma_{\rho\sigma} \mathcal{D}^2\right)  {d\sigma\over dy} . \nonumber \\ &&
 \eea{34a}
 We set $L/z_{IR}=0$ in the converging integrals and used the Bianchi identity $\mathcal{D}^{\mu} G_{\mn}(\gamma)=0$.
 To deal with the last integral in $y$ we use
 \bea
 \mathcal{D}_\mu V_\nu + \mathcal{D}_\nu V_\mu &=& \gamma^{\rho\sigma}V_\rho \partial_\sigma \gamma_{\mn} +
   \partial_\mu(\gamma^{\sigma\rho}V_\sigma) \gamma_{\rho\nu}+ \partial_\nu(\gamma^{\sigma\rho}V_\sigma) \gamma_{\rho\mu} 
   \nonumber \\ 
   &=& \n_\mu V_\nu + \n_\nu V_\mu + \Omega^{-2} V^\lambda(\partial_\lambda \Omega^2) g_{\mn} + \Omega^{2}\partial_\mu(\Omega^{-2})V_\nu +
     \Omega^{2}\partial_\nu(\Omega^{-2})V_\mu \nonumber \\ &=&
       \Omega^{2} \left[\n_\mu (\Omega^{-2} V_\nu) +  \n_\nu (\Omega^{-2} V_\mu) - V^\lambda(\partial_\lambda \Omega^{-2})g_{\mn}\right] ,
\eea{34b}
where indices are raised and lowered with the metric $g_{\mn}$. For $V_\mu=\partial_\mu(d\sigma/dy)$ eq.~(\ref{34b}) gives
\beq
\left( \mathcal{D}_{(\mu} \partial_{\nu)} - \gamma_{\mn} \gamma^{\rho\sigma} \mathcal{D}_\rho \partial_\sigma \right) {d\sigma \over dy} =
{d\over dy}\left[ \left( \n_\mu \partial_\nu -g_{\mn} \n^2\right)\sigma  -\partial_\mu \sigma \partial_\nu \sigma -{1\over 2} g_{\mn} \partial_\rho \sigma \n^\rho \sigma \right].
\label{34c}
\eeq
Inserting this identity in~(\ref{34a}) we find that the integral in $y$ is a total derivative
\bea
&& c \int_0^1 dy \int d^4 x \sqrt{-g}g^{\mu\rho}g^{\nu\sigma} \partial_\rho\sigma \partial_\sigma \sigma 
{d\over dy}\left[ \left( \n_\mu \partial_\nu -g_{\mn} \n^2\right)\sigma  -\partial_\mu \sigma \partial_\nu \sigma -{1\over 2} g_{\mn} 
\partial_\lambda \sigma \n^\lambda \sigma \right]. \nonumber \\ &=&
c \int d^4 x \sqrt{-g} \left( -\partial_\mu \tau \partial_\nu \tau \n^\mu \n^\nu \tau - {3\over 4}\partial_\mu \tau \n^\mu \tau \partial_\nu \tau \n^\nu \tau
\right).
\eea{34d}
 
The other terms in~(\ref{32}) are treated analogously. The cross term is 
\bea
&& -{c\over 4} \int_{L/z_{IR}}^1 dy \int d^4 x \sqrt{-g} \Big({d\over dy} (y^2 \Omega^{-2}) \Big) \Big( G_{\mn}(\gamma) C^{\mn} -
{1\over 3} G(\gamma) C\Big) \nonumber  \\ &=& -{c\over 4}  \int d^4 x \sqrt{-g} e^{2\tau}  \Big( G_{\mn}(e^{-2\tau} g) C^{\mn} -
{1\over 3} G(e^{-2\tau} g) C\Big) \nonumber \\ &&
-{c\over 2} \int_{L/z_{IR}}^1 dy \int d^4 x \sqrt{-g} (y^2 \Omega^{-2}) \Big(\mathcal{D}_\mu\mathcal{D}_\nu {d\sigma \over dy} 
\Big)C^{\mn} .
\eea{35}
Here too the indices are raised and lowered with the metric $g_{\mn}$ and the last term vanishes because $C_{\mn}$ is transverse w.r.t
the covariant derivative $\n_\mu$ and the traceless part of eq.~(\ref{34b}) is 
$ \Omega^{-2} \mathcal{D}_{(\mu} V_{n)_T} =\n_{(\mu} \Omega^{-2} V_{\nu)_T}$.

The last term is
\beq
c \int_{L/z_{IR}}^1 dy \int d^4 x \sqrt{-g} \Big({d\over dy}(y^4 \Omega^{-4})\Big) \Big(C_{\mn}^2 -{1\over 3} C^2 \Big) =
c  \int d^4 x \sqrt{-g} e^{4\tau} \Big(C_{\mn}^2 -{1\over 3} C^2 \Big) .
\label{36}
\eeq
Notice that~(\ref{35},\ref{36}) give only Weyl-invariant terms so the anomaly arises only from~(\ref{34})
\bea
S_{Anomaly} &=&  {c\over 4}  \int d^4 x  \sqrt{-g}\Big[ -\tau \Big( -G_{\mn}^2(e^{-2\tau} g)  + {1\over 3} G^2(e^{-2\tau} g) \Big)
+2\partial_\mu\tau \partial_\nu \tau G^{\mn}(e^{2\tau} g) + \nonumber \\ &&
-4\partial_\mu \tau \partial_\nu \tau \n^\mu \n^\nu \tau - 3\partial_\mu \tau \n^\mu \tau \partial_\nu \tau \n^\nu \tau \Big] .
\eea{37}
This is the same as eq.~(\ref{SA0}) with $a=c'$ once the identities
 \beq
 \int d^4x \partial_\mu \tau \partial_\nu \tau \n^\mu \n^\nu \tau=-{1\over 2} \int d^4 x (\n \tau)^2 \n^2 \tau, \quad
 \int d^4 x (\n \tau)^2 \n^2 \tau= \int d^4 x (\bar{\n} \tau)^2 [\bar{\n}^2\tau + 2( \bar{\n}\tau)^2]
 \label{iden}
 \eeq
are taken into account.

\section{Holography and Ward Identities}\label{holo}

The action~(\ref{5Daction}) computed on shell at a fixed value of the metric at $z=0$  has 
a dual holographic interpretation as
the effective action obtained from integrating out the CFT degrees of freedom~\cite{ahpr}. 
Because it is a 5D gravitational action, it is  invariant under the infinitesimal coordinate 
change $z=w + \omega(q)(w+L)$, $x^\mu= q^\mu + F^\mu(q,w)$, $F^\mu(q,0)=0$. When
$(w+L) \partial_\mu\omega(q)+ g_{\mu\nu}\partial_w F^\nu(q,w)=0$  the metric still has the form~(\ref{metric},\ref{Gexp}) 
and its boundary value transforms to
$g_{\mu\nu}(x)=(1-2\omega(x))g_{\mu\nu}(x)$. To first order in $\omega$, the limits of integration in~(\ref{5Daction}), 
$0\leq z \leq z_{IR}$, change to $-\omega(x)L\leq w\leq z_{IR}(x) -\omega(x)(z_{IR}(x)+L)$. 
 General coordinate invariance thus gives the equation
 \bea
 S[(1-2\omega(x)) g_{\mu\nu}(x), -\omega(x))L, z_{IR}(x) -\omega(x)(z_{IR}(x)+L)] &=&
 S[g_{\mu\nu}(x), 0, z_{IR}(x)] + O(\omega^2). \nonumber \\ &&
 \eea{6}
 This action diverges in the limit $L\rightarrow 0$. It can be written as~\cite{HenningsonSkenderis}
 \beq
 S[g_{\mu\nu},L,z_{IR}]= S_D[g_{\mn\nu}, L] +  S_F[g_{\mu\nu},z_{IR}] + O(L).
 \label{7}
 \eeq
 The divergent part depends only on the behavior of the metric near the $z=0$ boundary so it does not depend on 
 $z_{IR}(x)$~\cite{HenningsonSkenderis,hss,RZ}. The finite part does not depend on $L$ and the rest vanishes in the limit 
 $L\rightarrow 0$. The first variation of $S$ w.r.t. $\omega$ vanishes thanks to eq.~\eqref{6} so, up to terms
 that vanish in the $L\rightarrow 0$ limit we find
 \beq
 0= \delta_\omega S_D + \int d^4 x \sqrt{-g}\left[ \delta_\omega g_{\mu\nu} (x) {\delta S_F \over \delta g_{\mu\nu}(x)}
   + \delta_\omega z_{IR}(x) {\delta S_F \over \delta z_{IR}(x)}\right] .
   \label{8}
 \eeq
 Ref.~\cite{HenningsonSkenderis} finds ${ \delta S_D  / \delta \omega(x)} = -a(E-W^2)$ [see eq.~(\ref{anom0})].
 Notice that
 \beq
 \delta_\omega S_D = \int d^4 x \sqrt{-g}\left[ \delta_\omega g_{\mu\nu} (x) {\delta S_D \over \delta g_{\mu\nu}(x)}
   + \delta_\omega L(x) {\delta S_D \over \delta L(x)}\right].
 \eeq
 The second term is a variation w.r.t. $L$. In RS1 $L$ is the position of the UV brane, which is kept
 fixed. Keeping $L$ fixed and introducing a bare Einstein-Hilbert term on the UV brane explicitly breaks Weyl
 invariance. The terms that break Weyl invariance in $S_D$ are a 4D cosmological constant
 $\propto L^{-4} \int d^4x \sqrt{-g}$ and an Einstein-Hilbert term $\propto L^{-2} \int d^4x \sqrt{-g}R(g)$.
 The holographic interpretation of these terms is that they are induced by the CFT loops. The induced 4D cosmological constant is 
 canceled by the brane tension while the bare and induced Newton constants give the 4D Newton constant $G=2/M^2$. 
 So, $S$ satisfies the equation
 \bea
 && \int d^4 x \sqrt{-g}\left[ \delta_\omega g_{\mu\nu} (x) {\delta S \over \delta g_{\mu\nu}(x)}
   + \delta_\omega z_{IR}(x) {\delta S \over \delta z_{IR}(x)}\right] = 
   - \int d^4 x \sqrt{-g}\left[ \delta_\omega L(x) {\delta S_D \over \delta L(x)}\right]  +O(L) \nonumber \\ &=&
   \int d^4 x \sqrt{-g}\omega(x)   a(E-W^2)  + 
  \int d^4 x \sqrt{-g}\delta_\omega g_{\mu\nu} (x) {\delta S_D \over \delta g_{\mu\nu}(x)}  +O(L) .\nonumber \\ &=&
  \int d^4 x \sqrt{-g}\omega(x) a(E-W^2)    - {2\over 16\pi G} 
  \int d^4 x \sqrt{-g}\omega(x)  R(g) + O(L)    .
  \eea{10a}
This is indeed the Ward identity for a spontaneously broken CFT coupled to gravity.

\section{Derivation of the sub-leading terms}\label{a-c}

\subsection{Subleading $W^2$ contributions to the anomaly}
The general anomaly is the sum of the term $R_{\mn}^2 -{1\over 3}R^2$ and the square of the Weyl tensor.
The additional 5D term that gives rise to this anomaly is~\cite{sub1,sub2,sub3}
\beq
S_W= \alpha k\int_L^{L+z_{IR}} dz \int d^4x \sqrt{G} C^M_{NPQ} C_M^{NPQ} ,
\label{38}
\eeq
where $C^M_{NPQ}(G)$ is the Weyl tensor in 5D, $\alpha\ll c$, and for convenience in this section 
we will use the $z$ coordinate that is shifted by $L$ with respect  to the one used in the 
previous sections, so the $z$ integration range is, $L\leq z\leq L+z_{IR}\simeq z_{IR}$ .
Its key property is that it is Weyl Invariant: 
$C^M_{NPQ}(\omega^2 G)=C^M_{NPQ}(G)$. We only need to evaluate this integral on the zero-mode metric because
the contribution of $h_{\mn}$ terms is $O(z_{IR}^2 \partial_\mu^2)$. 
By defining $\Omega$ as in section 4.1 we can write this metric as
\beq
ds^2={\Omega^2L^2\over z^2} (\Omega^{-2} (1-z\partial_z\Omega \Omega^{-1} )^2 dz^2 + g_{\mn} dx^\mu  dx^\nu), \quad \mathrm{with}\; {dg_{\mn}\over dz}=0\,.
\label{39}
\eeq
The coordinate change $T=z\Omega^{-1} $, 
$(\Omega^{-1}-z\partial_z \Omega \Omega^{-2} )dz = dT +T\partial_\mu\Omega \Omega^{-1}dx^m$ transforms metric~(\ref{39}) 
into
\bea
ds^2 &=& {L^2\over T^2}d\tilde{s}^2={L^2\over T^2}\left[dT^2 +2T\partial_\mu\sigma dx^m dT+ 
\left(g_{\mn}(x) +T^2\partial_\mu\sigma \partial_\nu\sigma\right)dx^m dx^\nu \right], \quad \sigma=\log \Omega. \nonumber \\ &&
\eea{40}
Since the Weyl tensor is invariant under conformal rescaling, we can compute it on the metric $d\tilde{s}^2$. Moreover, since
$T\lesssim z_{IR}$, when we compute curvature tensor in the effective 4D theory we expand in powers of $z_{IR}\partial_\mu$. So, in considering the most relevant terms in the expansion, we can neglect all terms proportional to $T$ in the 
metric {\em except} for
those where derivatives of $T$ cancel out the $T$ dependence. In the Riemann tensor there is only one such term:
\beq
R_{T\mu T\nu}= -\partial_T K_{\mn} = -{1\over 2} \partial_T \left( 2T\partial_\mu\sigma \partial_\nu\sigma -2  T \n_\mu\partial_\nu \sigma\right)
= -\partial_\mu\sigma\partial_\nu \sigma + \n_\mu\partial_\nu \sigma.
\label{41}
\eeq
Hence $R_{\mu\nu\rho\sigma}=R_{\mu\nu\rho\sigma}(g)$ and
\bea
R_{\mn}&=&R_{\mn}(g) - \partial_\mu\sigma\partial_\nu \sigma + \n_\mu\partial_\nu \sigma, \quad R_{TT} = -\partial_\mu\sigma \n^\mu \sigma + \n^2 \sigma , \nonumber \\  R&=& R(g) - 2\partial_\mu\sigma \n^\mu \sigma + 2\n^2 \sigma. 
\eea{42}
A short calculations gives 
\beq
C^M_{NPQ} C_M^{NPQ} =   4R_{T\mu T\nu}R^{T\mu T\nu} +R_{\mu\nu\rho\sigma}R^{\mu\nu\rho\sigma} -{4\over 3} R_{\mn} R^{\mn} -{4\over 3} R_{TT}R^{TT}+{1\over 6} R^2 ,
\label{43}
\eeq
with indices $\mu,\nu,..$ raised and lowered with the metric $g_{\mn}$. Substituting eqs~(\ref{41},\ref{42}) into~\eqref{43} a tedious but 
straightforward calculation gives
\bea
\sqrt{g}C^M_{NPQ} C_M^{NPQ} &=& \sqrt{g}\left[R_{\mu\nu\rho\sigma}(g)R^{\mu\nu\rho\sigma}(g) -2R_{\mn}(g)R^{\mn}(g) + 
{1\over 3} R^2(g)\right]  \nonumber \\
&& +\sqrt{\gamma}\left[ {2\over 3}R_{\mn}(\gamma) R^{\mn}(\gamma) -{1\over 6} R^2(\gamma) \right],
\eea{44}
where all the terms in the second bracket are defined w.r.t. the metric $\gamma_{\mn}=e^{2\sigma} g_{\mn}$, which is also used there to raise 
and lower indices.
Notice that the first term is proportional to the Weyl anomaly term 
$W^2=  \left(R_{\mu\nu\rho\sigma}R^{\mu\nu\rho\sigma} -2 R_{\mn} R^{\mn} + {1\over 3} R^2 \right)$. We can therefore recall the definition
$\tau=-\sigma |_{y=1}$ and rewrite $S_W$ as
\bea
S_W &=&  \alpha\int d^4x \int_L^{z_{IR} e^{\tau}} {dT\over T} \left[\sqrt{g} W^2 + \sqrt{\gamma}\left( {2\over 3}R_{\mn}(\gamma) R^{\mn}(\gamma) -{1\over 6} R^2(\gamma) \right)\right]  \nonumber \\
&=& 64\alpha\int d^4x \sqrt{g} \tau W^2 + \alpha \int d^4x \int_L^{z_{IR} e^{\tau}} {dT\over T} \sqrt{\gamma}\left( {2\over 3}R_{\mn}(\gamma) R^{\mn}(\gamma) -{1\over 6} R^2(\gamma) \right). \nonumber \\ &&
\eea{45}
The first term is the Weyl tensor contribution to the anomaly. To compute the
last term we transform back from $T$ to $z$ using
$dT= (\Omega^{-1}-z\Omega^{-2}\partial_z \Omega)dz - z\Omega^{-2} \partial_\mu \Omega dx^\mu$. Setting $z=z_{IR} y$, the second term in~\eqref{45} becomes
\beq
    \alpha \int_{L/z_{IR}}^1 {dy\over y} \int d^4x  \sqrt{\gamma}N \left( {2\over 3}
    \gamma^{\mu\rho}\gamma^{\nu\sigma}G_{\mn}(\gamma) G^{\rho\sigma}(\gamma) -{1\over 6} [\gamma^{\mn}G_{\mn}(\gamma)]^2 \right).
\label{46}
\eeq
This action vanishes when  $G_{\mn}(\gamma)=0$ so it can be canceled up by a redefinition of the metric $h_{\mn}$ in eq.~(\ref{25}). Concretely, we set
\beq
h_{\mn} = \bar{h}_{\mn}+ \Delta h_{\mn}=
\bar{h}_{\mn} + {z_{IR}^2y^2  \over \Omega^2 N}  \left[ V G_{\mn}(\gamma) +B \gamma_{\mn} G(\gamma) \right],
\label{47}
\eeq
where $\bar{h}_{\mn}$ solves eq.~(\ref{27}). Substituting in~(\ref{25}) the linear term in $\Delta h_{\mn}$ vanishes
and, discarding terms of higher order in $z_{IR}$ we get the additional $O(\Delta h_{\mn})^2$ term
\bea
&& {c\over {z_{IR}}^4} \int_{L/z_{IR}}^1 {dy\over y^3} \int d^4 x  \sqrt{-g} 
 {\Omega^4\over 4N} \left( {d\Delta h \over dy}{d\Delta h\over dy}- {d\Delta h^{\mn} \over dy}{d \Delta h_{\mn}\over dy}  \right) 
 \nonumber \\ &=&
 c \int_{L/z_{IR}}^1 {dy\over y^3} \int d^4 x  \sqrt{-\gamma} \left[ V^2 G_{\mn}(\gamma)^2  - (V^2+6VB+12B^2) G(\gamma)^2 \right].
 \eea{48}
 Hence $V= \pm{\alpha\over c} \sqrt{6}/3$, $B= \mp{\alpha\over c} \sqrt{6}/12$ so we need $\alpha>0$. When translated into the 
 coefficients of the general anomaly,  $\langle T^\mu_\mu \rangle =c' W^2 -a E$,  it means $c'>a$. Presumably the other sign could be dealt with by studying the
 5D Euler density (Lovelock) action as in ref.~\cite{sub3}.
 
 We note that the departure from the $a=c'$ can also be achieved  by introducing
the bulk 5D Gauss-Bonnet term \cite {sub2}. In this case, similarly to the case of the 5D Weyl-tensor square 
considered above,   one would be introducing in the bulk another scale below $M_5$ at which the 
5D theory would become strongly coupled (or else these 5D terms would have to be suppressed by 
$M_5$, in which case such terms can't be differentiated from generic 
terms emerging from quantum gravity at $M_5$.)  

In the next section we will discuss a framework in a weakly coupled theory that enables to relax the condition $a=c'$.
 
\subsection{Sub-leading terms from a boundary-localized QFT}

There is another way  for the theory to depart from the $a=c'$ limit. 
Suppose there  is a 4D boundary QFT localized on a positive tension brane. 
This QFT is coupled to the metric ${\hat g}_{\mn}(x,z=0)=g_{\mu\nu}(x)$. Let us integrate out the boundary 
QFT in the path integral,  and use dimensional regularization to deal with the 
divergences. This would make the brane world-volume to become $D=4-2\delta$ dimensional, with $\delta \to 0$ 
to be taken after all the divergences are subtracted. The result of this calculation --
with the massless limit taken at the very end -- has long 
been known (see, Duff's work in \cite {Duff} and an overview in \cite {BirDav}), it is proportional to 
\beq 
\Gamma (2-D/2) \,\int d^Dx\,\sqrt {-g}  \(a_b E(g) -c_b ( {R}_{\mu\nu\alpha\beta}^2 - 2 {R}_{\mu\nu}^2+{R}^2/3 ) \)\,,
\label{Wdimreg}
\eeq
where $\Gamma$ is the Euler gamma function, $E$ is the Euler (Gauss-Bonnet) 
invariant, ${E}={R}_{\mu\nu\alpha\beta}^2 - 4 {R}_{\mu\nu}^2+{R}^2$, $a_b$ and $c_b$ are calculable 
coefficients, and in general $a_b\neq c_b$.
The combination of the curvature  square invariants proportional to $c_b$ combines into the Weyl tensor 
squared only after the $\delta \to 0$ limit is taken.  Since the term (\ref {Wdimreg}) is localized on the UV brane, 
it will have to be added to the holographic 4D action. This would shift the holographic theory 
away from the $a=c'$ limit, to $a_{tot}\neq c_{tot}$, where $a_{tot}=a+a_b$ and $c_{tot}=c'+c_b$. 
For simplicity, we  will focus on the $E$ term in (\ref {Wdimreg}), 
while requiring a QFT that has $a_b$ not equal to  $c_b$.

As we mentioned earlier, it is known
that the trace of the variation of (\ref {Wdimreg})  
with respect to $g$ gives the right trace anomaly equation after taking the limit $\delta \to 0$. 
If so, then  (\ref {Wdimreg}) should also contain the Riegert action in the $\delta \to 0$ limit.
This can be shown under the assumption of validity of the 
analytic continuation  to negative values of $\delta$,  and by representing the 
interval of  the $4+n$ dimensional theory as, 
\beq
d_{4+n}s = g_{\mu\nu}(x) dx^\mu dx^\nu + e^{2\tau(x)} d^2_n z\,,
\label{dimreginterval}
\eeq
where $n=-2\delta$.  Then, taking the limit $n=-2 \delta \to 0$,  the divergent coefficient proportional 
to $1/\delta$ coming from the $\Gamma$ function is cancelled by the terms proportional 
to $\delta$ coming from $E$ and the resulting finite term  is exactly the Riegert term  
written in terms of  the metric $g$ and the scalar $\tau$ \cite {LuPang} (see, also 
\cite {Kurt} and references therein)
\beq
\int d^4x \sqrt {- {g} } \left (  \tau { E} + 4 {G}^{\mu\nu} {\nabla}_\mu  \tau {\nabla}_\nu  \tau  - 4 ({\nabla}^2  \tau) ({\nabla}  \tau)^2 + 
2 ({ \nabla}  \tau )^4\right)\,.
\label{Riegert_gtau}
\eeq
Furthermore, using $g= e^{2\tau}{\bar g}$ in (\ref {Riegert_gtau}), one recovers the $a$-terms of  (\ref {SA0}).

To summarize, with the help of a localized QFT on the positive tension RS brane
one obtains a weakly coupled completion 
for the trace anomaly action with $a_{tot}\neq c_{tot}$.

\section{Discussions}\label{disc}

The location of the IR brane in the 5D theory considered in the present work is not stabilized. One could introduce 
a potential  to stabilize its location in the 5th dimension, and stabilize the radion via the Goldberger-Wise (GW) mechanism 
\cite {gw}; however, this would in general alter the ability of the resulting  
theory to correctly recover the 4D trace anomaly equation. That said, if the scales in the GW potential
are much smaller than $\bar M$, the resulting equation could give a good approximation to the 
trace anomaly equation,  \cite {KS}.  Without the GW mechanism, 
the vacuum expectation value of the $\Phi$ field is a modulus. This VEV sets the value of the scale $\bar M$, 
which is not dynamically determined. As we discussed,  $\bar M$ has to be significantly below the Planck scale for the 
trace anomaly effective  field theory to be  distinguished from other higher dimensional terms, 
which are suppressed by the Planck scale. 
Furthermore, $1/{\bar M}$ serves as a constant determining the self-interactions of the radion, which in 
the 4D holographic theory can be regarded as a dilaton of spontaneously broken conformal symmetry.

What are the couplings of the matter fields to the metric and dilaton? The answer depends on where the matter fields are 
assumed to be placed in 5D. If they are localized on the IR brane, as they are in the RS1 scenario for the sake of solving 
the hierarchy problem, 
then matter would coupe to the metric times $\Omega^4$.   Because of the presence of a long-range 
radion such a theory would be ruled out observationally in our case.  
In the holographic approach adopted in this work it is 
more natural to assume that matter  --that is the weakly coupled matter that we have added to the strongly coupled CFT--
 is localized on the UV brane, in which case 
it would couple to the metric $g$. Based on symmetry considerations 
 matter was coupled in \cite {GG} to the metric ${\hat g} =g(1-\Phi^2/M^2)$.
Since $M=M_{\rm Pl}/\sqrt{2}\gg{\bar M}$, the difference between coupling to $g$ and $\hat g$ is small. 

In either case, there is a fifth force produced by the long-range dilaton. 
If the matter couples to $g$  the coupling to the dilaton appears at the linear level and its strength 
is proportional to $(\bar M/M)^2$, while in the case when matter couples to ${\hat g}$ the coupling  to dilaton 
only emerges at the nonlinear level and its strength is proportional to $a\,(\bar M/M)^2$, as shown by 
Tsujikawa \cite {Tsujikawa} --who has recently found a black hole solution  in the effective trace anomaly 
action and explicitly calculated the corrections to the Schwarzschield geometry proportional 
to $a\,({\bar M}/M)^2$. Comparison of these predictions with observational data 
will likely impose strong bounds on the value of the scale $\bar M$.

While the 5D construction serves the point of identifying a completion of the 4D effective 
theory above its strong scale $\bar M$, the 4D effective theory should still be a 
more convenient tool for practical calculations:  in general, it is significantly easier to work with 
4D  differential equations  (in many symmetric cases being ordinary differential equations)
rather than to work with 5D partial differential equations.

\vspace{0.5in}

{\bf Acknowledgments:} We'd like to thank Justin Khoury, Shinji Tsujikawa,  and Giorgi Tukhashvili 
for useful discussions. M.P. is supported in part by the Leverhulme Trust through a Leverhulme Visiting Professorship.  This work is supported in part by NSF grant PHY-2210349.

\vspace {0.5in}

%\bibliographystyle{unsrt.bst}
%\bibliography{RSScalar}

%\bibliographystyle{utphys}
%\bibliography{bibliography} 

\end{document}